\documentclass{aa}   
\usepackage{CJKutf8}
\usepackage{graphicx}
\usepackage[breaklinks,colorlinks,citecolor=blue,linkcolor=blue]{hyperref}
\usepackage{txfonts}
\usepackage{lscape}
\usepackage{epstopdf}
\usepackage{CJK}
\usepackage{float}
\usepackage[absolute]{textpos}
\usepackage{amsmath}
\usepackage{mathrsfs}

\usepackage{arydshln}

\begin{document} 

   \title{Shared star formation in the Milky Way and  Magellanic Clouds}


%
%
%

\author{Xunchuan Liu \begin{CJK}{UTF8}{gbsn}(刘训川)\end{CJK}$^1$\thanks{liuxunchuan001@gmail.com} \and Yu Cheng$^2$}

   \institute{Leiden Observatory, Leiden University, P.O. Box 9513, 2300RA Leiden, The Netherlands \and 
   National Astronomical Observatory of Japan, 2-21-1 Osawa, Mitaka, Tokyo, 181-8588, Japan
   }

\abstract{We investigate the structural and evolutionary similarities between star formation patterns in different environments by comparing the dense clump populations in the Milky Way (MW) from the ATLASGAL survey with those from the \textit{Herschel} HERITAGE survey in the Magellanic Clouds (MCs). Our analysis reveals that MW and MC clumps behave as physical analogs, sharing consistent dust temperature distributions, mass spectra, and luminosity evolutionary trends. We establish that the warmest MC clumps and the most distant MW clumps share an identical fiducial parent structure bounded by a natural spatial scale of $\sim 1$~parsec, serving as the direct precursors to open clusters. Closer MW clumps are resolved into discrete sub-clumps, whereas colder MC clumps suffer from peripheral envelope mass blending. Furthermore, the global spatial layout of clumps in the LMC and the MW shares a remarkably similar pattern when adjusting for galaxy size, suggesting a nested, hierarchical distribution. The clump-based star formation rates are calibrated to be $\sim 0.4~M_\odot\,\rm yr^{-1}$ for the LMC and $\sim 0.1~M_\odot\,\rm yr^{-1}$ for the SMC, confirming that the LMC is currently experiencing an active, ongoing star formation burst captured within a short ($< 10^6$~yr) snapshot timescale.}


   \maketitle
    \nolinenumbers

\section{Introduction} \label{sec:intro}
Massive stars dominate the evolution of galaxies, with their birth occurring as clusters within dense gas clumps that are considered the fundamental building blocks sharing intrinsic characteristics across galaxies \citep{2003ApJ...585..850M, 2004MNRAS.349..735B, 2024RAA....24b5009L, 2025RAA....25b5020L}. Consequently, it is possible to investigate the ongoing galactic star formation rate (SFR) by measuring the flux of dense gas \citep[i.e., the star formation law;][]{1998ApJ...498..541K, 2004ApJ...606..271G} or, more directly, by counting the number of star-forming clumps when observational resolution permits.

To render the clump-counting method applicable, we must quantitatively describe the evolutionary patterns of dense clumps within the Milky Way (MW), thereby providing a baseline template for comparisons with external galaxies. Recently, \citet{2025arXiv251025436L} introduced a statistical framework utilizing the cumulative distribution function (CDF) of dust temperature to map static clump observations onto a linear timeline. This approach was inspired by \citet{2024RAA....24g5001L}, who employed the CDF of the infrared spectral energy distribution (SED) slope to trace the evolutionary timescales of protoplanetary disks. By applying this method to the ATLASGAL \citep{2009A&A...504..415S} and ALMAGAL \citep{2025A&A...696A.149M} surveys, \citet{2025arXiv251025436L} discovered that the virial masses of Galactic clumps \citep[e.g.,][]{2018MNRAS.473.1059U} increase exponentially over time, establishing a distinct signature of accelerating star formation. Within this framework, the rapid increases in clump luminosity \citep{2009A&A...504..415S} and maximum core mass \citep{2025A&A...696A.151C}, driven by emerging massive young stars and vigorous protostellar accretion, align well with this accelerating mass growth.

The Magellanic Clouds (MCs), including the Large Magellanic Cloud (LMC) and Small Magellanic Cloud (SMC), provide an ideal laboratory to test whether this evolutionary pattern is universal. The LMC is our nearest face-on dwarf galaxy, located at a distance of $D\sim 50$~kpc \citep{2019Natur.567..200P}. The SMC is an elongated, irregular dwarf galaxy situated at a slightly larger distance of $D\sim 62$~kpc \citep{2020ApJ...904...13G}. They feature a sub-solar metallicity environment \citep[$\sim 0.5~Z_{\odot}$ for the LMC, and $\sim 0.2~Z_{\odot}$ for the SMC;][]{2020MNRAS.497.3746C, 2021MNRAS.507.4752C}, making them typical analogues of main-sequence galaxies at Cosmic Noon \citep[$z \sim 1\text{--}3$,][]{2026ApJ..1004L..31R}. Far-infrared observations toward the LMC and SMC are available from the \textit{Herschel} HERITAGE survey \citep{2013AJ....146...62M, 2014AJ....148..124S}, which mapped the dust continuum across the Magellanic Clouds at the PACS bands (100 and 160~$\mu$m) and the SPIRE bands (250, 350, and 500~$\mu$m). In the PACS 100~$\mu$m band, \textit{Herschel} yields its highest angular resolution of $\sim 7''$ \citep{2013A&A...553A.132M}, corresponding to a physical spatial resolution of approximately 1.7~pc (2.1~pc) at the distance of the LMC (SMC). This scale is highly comparable to the physical resolution of $\sim 1$~pc achieved by the ATLASGAL survey's $19.2''$ beam \citep{2009A&A...504..415S} toward typical dense clumps located 10~kpc away in the Galactic plane. 

This closely matched physical resolution enables us to directly evaluate whether high-mass star formation follows the same global patterns across these systems by comparing their dust temperature distributions \citep[as a tracer of evolution;][]{2025arXiv251025436L}, luminosities, masses, and spatial arrangements. Such a cross-galaxy comparison allows us to test the invariance of this evolutionary framework against the significant differences in metallicity and radiation environments that exist between the MCs and the MW. Benefiting from its nearly face-on geometry and more extended features compared to the SMC, the LMC serves as the most promising target for obtaining rigorous quantitative comparisons. Furthermore, calibrating this clump-counting framework naturally provides an independent measurement of the ongoing global SFR in the MCs.

This paper is organized as follows. In Sect.~\ref{sec:data}, we describe the sample of MC and MW clumps utilized in this work. In Sect.~\ref{sec:calibration}, we statistically compare the clump parameters to justify the similarity between the observed MC and MW clumps. In Sect.~\ref{sec:lmc_sfr}, we calibrate the clump masses across different samples based on our finding of a fiducial 1~parsec clump scale, and then estimate the global star formation rates of the MW and MCs. Finally, we present our discussion in Sect.~\ref{sec:discu}, followed by a brief summary in Sect.~\ref{sec:summary}.

\section{Sample} \label{sec:data}

\subsection{Milky Way clumps} \label{subsec:galactic_clumps}
To align the MC clumps with the MW baseline, we incorporate the CDF timeline framework \citep{2025arXiv251025436L} established for the MW clumps. The ATLASGAL survey provides an unbiased and complete sample of approximately 8\,000 dense MW clumps at 870~$\mu$m, covering 420~deg$^2$ of the inner MW plane within $-80^\circ < \ell < 60^\circ$ and $|b| < 1.5^\circ$ \citep{2009A&A...504..415S, 2018MNRAS.473.1059U}. We quote the clump masses, dust temperatures, luminosities, and distances from the catalog \citep{2018MNRAS.473.1059U}. These clumps possess well-probed dust temperatures ($T_{\rm dust}$) ranging from 7.9 to 56.1~K. Under the fundamental assumption that this sample is unbiased and complete within its sky coverage, the empirical CDF of the $T_{\rm dust}$ distribution has been directly linked via a linear relation to the physical evolutionary time $t$ in \citet{2025arXiv251025436L}. This non-parametric mapping linearizes $T_{\rm dust}$ into a temporal coordinate normalized by the typical clump lifetime $t_0$, scaling the mapped timeline $t$ to span from 0 to $\sim 2$.

\subsection{Magellanic Cloud clumps} \label{sec_lmcclump}
\begin{figure}[!t]
    \centering
    \includegraphics[width=0.99\linewidth]{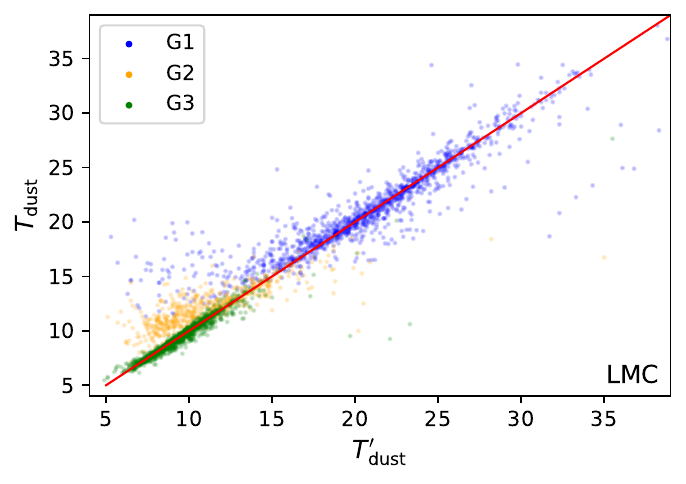}
    \includegraphics[width=0.99\linewidth]{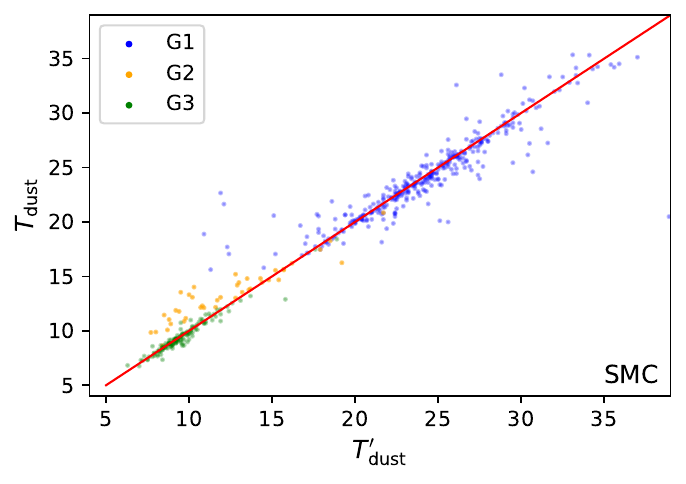}
    \caption{Comparison between the fitted dust temperatures ($T_{\rm dust}$; see Sect.~\ref{subsec:dust_temp}) of this work and the catalog values ($T_{\rm dust}'$) for the LMC (upper panel) and SMC (lower panel) clumps in three groups (see Sect.~\ref{sec_lmcclump}). The red solid line represents $y=x$.}
    \label{fig_TdustFit}
\end{figure}
To construct a complete, unbiased sample of MC clumps, we utilize the high-reliability point-source catalogs from the HERITAGE survey \citep{2013AJ....146...62M, 2014AJ....148..124S}. The original source catalog classifies detected far-infrared sources into various stellar, interstellar, and extragalactic populations, including background galaxies, evolved stars, planetary nebulae, compact supernova remnants, young stellar objects (YSOs), candidate dust clumps, and unclassified sources \citep{2014AJ....148..124S}. Among these broad classifications, the target YSO and dust clump categories are further broken down into four specific sub-labels based on their selection confidence: probable YSOs, possible YSOs, probable dust clumps, and possible dust clumps. Rather than separating these groups based on the presence of hot dust emission, we combine all four sub-labels into a single, comprehensive population. This unified treatment mirrors the approach used in MW surveys like ATLASGAL, where the entire physical sequence of dense clumps, including quiescent clumps, protostellar clumps, YSOs, and \ion{H}{ii} regions \citep{2018MNRAS.473.1059U, 2022MNRAS.510.3389U}, is retained together.

We further require detections in all three SPIRE bands (250, 350, and 500~$\mu$m), or across at least four of the five total \textit{Herschel} bands \citep{2010A&A...518L...1P}, ensuring that each source contains a substantial, cold dust mass reservoir rather than just an isolated, hot stellar envelope. This rigorous data requirement provides a clean and complete dataset, within the detection limits of \textit{Herschel}, to construct the evolutionary timeline of dense clumps. We divide the clumps into three subgroups based on their short-wavelength detection status: $G_1$ contains sources with 100~$\mu$m detections; $G_2$ contains sources lacking 100~$\mu$m detections but retaining 160~$\mu$m detections; and $G_3$ contains sources with neither 100 nor 160~$\mu$m detections. In the LMC, there are 963, 903, and 2\,216 clumps in $G_1$, $G_2$, and $G_3$, respectively. In the SMC, there are 422, 58, and 162 clumps in $G_1$, $G_2$, and $G_3$, respectively.

\begin{figure*}
    \centering
    \includegraphics[width=0.95\linewidth]{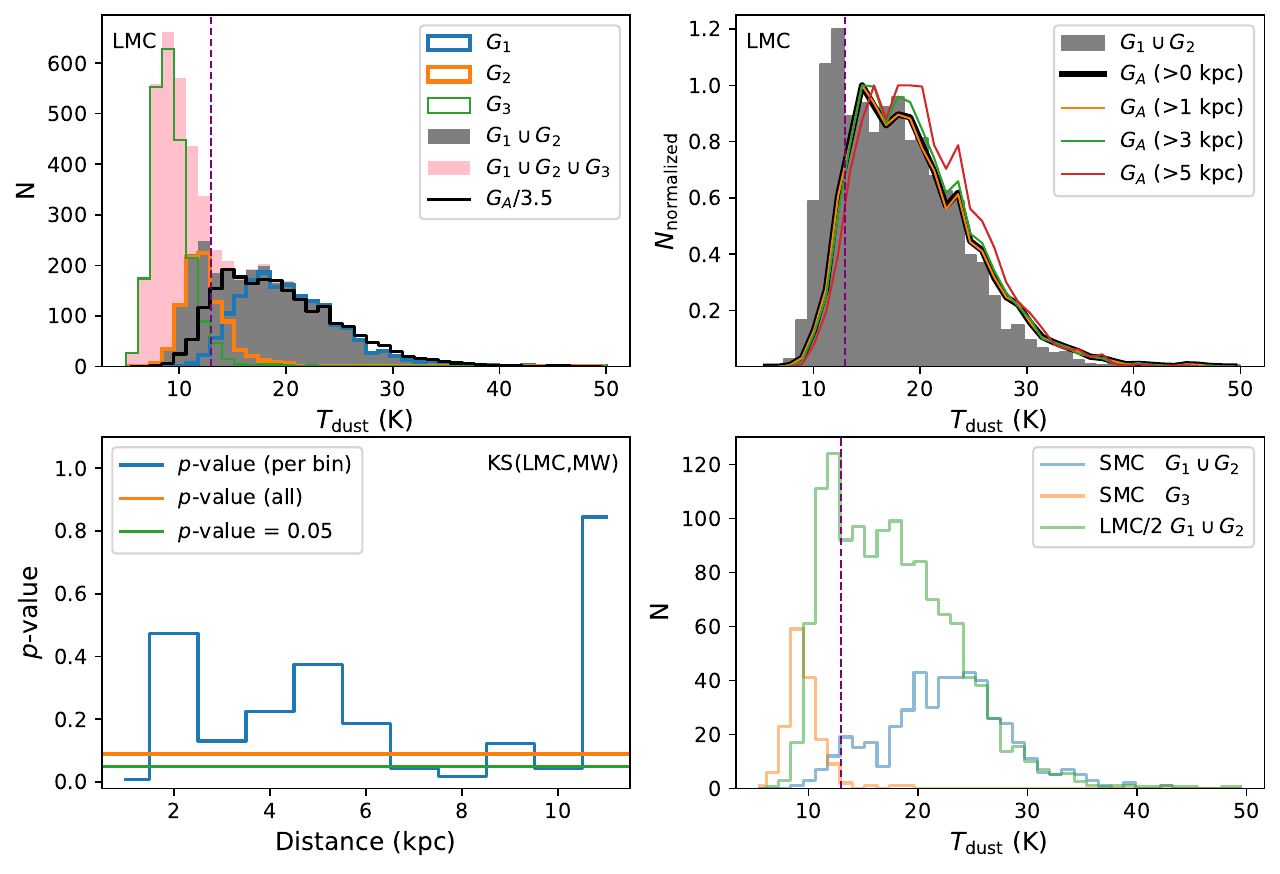}
    \caption{Upper left: Number distribution of $T_{\rm dust}$ for LMC clumps across different subgroups (Sect.~\ref{sec_lmcclump}) and ATLASGAL clumps (marked as $G_{\rm A}$), with the $G_{\rm A}$ distribution scaled down by a factor of 3.5. Upper right: Normalized number distributions of the combined subgroup $G_1 \cup G_2$ for the LMC, plotted alongside ATLASGAL clumps exceeding different distance thresholds (Sect.~\ref{sec_TdustDistr}). Lower left: The $p$-value of the Kolmogorov-Smirnov (K-S) test evaluated between the dust temperature distributions of LMC $G_1 \cup G_2$ and ATLASGAL samples within various distance bins. The horizontal green line indicates the $p=0.05$ significance threshold. Lower right: Comparison of the $T_{\rm dust}$ number distributions between LMC (scaled down by a factor of 2) and SMC clumps. The dashed vertical purple line in the upper panels and the lower right panel marks a dust temperature of 13~K.}
    \label{fig:TdustDistr}
\end{figure*}

\subsection{Dust temperature of Magellanic Cloud clumps} \label{subsec:dust_temp}
To derive dust temperatures ($T_{\rm dust}$) and dust masses ($M_{\rm dust}$) for the MC clumps, we perform an optically thin, single-temperature graybody fit with a fixed dust emissivity index of $\beta = 1.5$, aligning with the baseline assumptions of the original catalog paper \citep{2014AJ....148..124S}. Crucially, we do not apply upper limit constraints for undetected bands, fitting only those channels with confirmed detections. At short wavelengths, non-detections in the PACS channels may physically indicate deeply embedded structures that are not completely optically thin to their own compact heating source. Conversely, at long wavelengths, non-detections in the SPIRE 500~$\mu$m band are typically driven by beam-smearing and spatial confusion, given its coarser angular resolution of $\sim 36''$ \citep{2010A&A...518L...3G}. Specifically, the source flux density $F_{\nu}$ is modeled via the standard Planck blackbody function $B_{\nu}(T_{\rm dust})$ as:
\begin{equation}
    F_{\nu} = \frac{ \kappa_{\nu} M_{\rm dust} B_{\nu}(T_{\rm dust})}{D^2},
\end{equation}
where the dust opacity coefficient follows the MW baseline value $\kappa_{\nu} = \kappa_0 (\lambda/\lambda_0)^{-1.5}$ with $\kappa_0 = 1.85$~cm$^2$\,g$^{-1}$ at \mbox{$\lambda_0 = 300$~$\mu$m} \citep{1994A&A...291..943O}. Note that the choice of reference opacity $\kappa_0$ acts purely as an absolute normalization baseline that leaves the derived $T_{\rm dust}$ entirely unchanged, while the calculated $M_{\rm dust}$ scales inversely with it.

To derive the total clump gas mass ($M_{\rm c}$), we adopt:
\begin{equation}
    M_{\rm c} = \alpha_{\rm GDR} M_{\rm dust},
\end{equation}
where $\alpha_{\rm GDR}$ is the gas-to-dust mass ratio. We set $\alpha_{\rm GDR} = 200$ for the LMC and $\alpha_{\rm GDR} = 500$ for the SMC to account for their lower metallicities. Note that in the diffuse gas of the MCs, $\alpha_{\rm GDR}$ can be significantly higher, occasionally exceeding 1\,000 \citep{2009ApJ...690L..76G, 2011ApJ...737...12L}. However, in denser environments, $\alpha_{\rm GDR}$ decreases substantially due to dust grain growth and dust coagulation \citep[e.g.,][]{2014ApJ...797...86R}. Consequently, our adopted values represent the standard dense-phase MW baseline ($\alpha_{\rm GDR} \sim 100$) scaled inversely by the metallicities of the LMC and SMC relative to the MW.

A comparison between our fitted $T_{\rm dust}$ and the values ($T_{\rm dust}'$) presented in the catalog of \citet{2014AJ....148..124S} is shown in Figure~\ref{fig_TdustFit}. Overall, the two datasets match each other well across the entire sample population. For the sources in $G_1$ and $G_2$, our derived $T_{\rm dust}$ values tend to be slightly higher than the catalog values at the low-temperature end, making $T_{\rm dust}$ for $G_1$ and $G_2$ overall larger than 15~K and 10~K, respectively. This systematic shift effectively decouples the dust temperature distributions of the three groups, which are otherwise heavily overlapped, providing a clearer statistical separation between their evolutionary phases.

\subsection{Luminosity of Magellanic Cloud clumps}\label{sec_lumical}
We evaluate the total bolometric luminosities ($L_{\rm tot}$) of the MC clumps by combining their mid-infrared and far-infrared emission components ($L_{\rm tot} = L_{\rm MIR} + L_{\rm FIR}$). The far-infrared component ($L_{\rm FIR}$) is calculated by integrating our best-fit single-temperature graybody profile. For the mid-infrared component ($L_{\rm MIR}$), we compute the trapezoidal area under the spectral energy distribution (SED) across the detected \textit{Spitzer} channels spanning 3.6 to 24~$\mu$m. The mid-infrared luminosity is calculated explicitly as:
\begin{equation}
    L_{\rm MIR} = 4\pi D^2 \sum_{i=1}^{N-1} \frac{1}{2} \left( \nu_i F_{\nu, i} + \nu_{i+1} F_{\nu, i+1} \right) \ln\left(\left|\frac{\nu_{i+1}}{\nu_i}\right|\right),
\end{equation}
where $F_{\nu, i}$ is the observed flux at the $i\text{-th}$ \textit{Spitzer} channel, and $\nu$ is the frequency.

\begin{figure}[!t]
    \centering
    \includegraphics[width=0.99\linewidth]{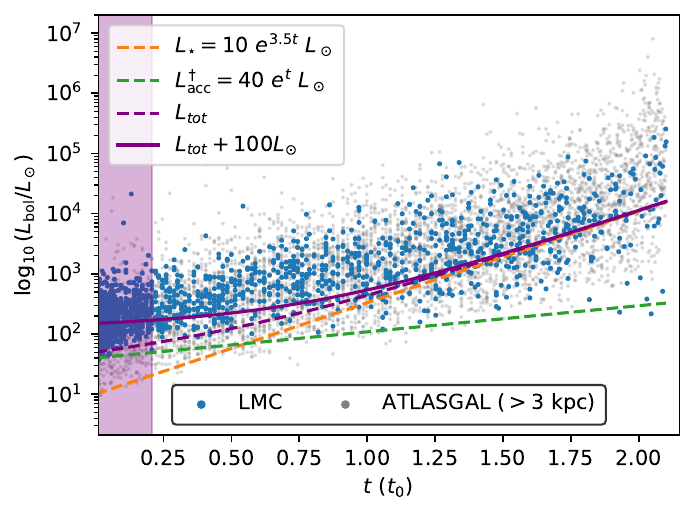}
    \includegraphics[width=0.99\linewidth]{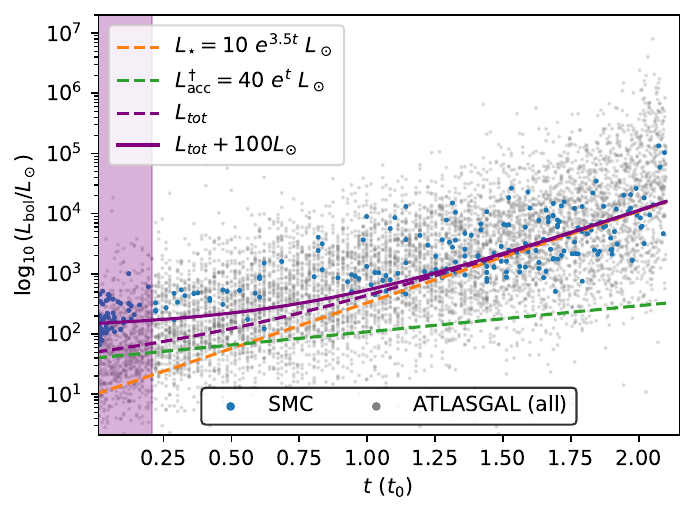}
    \caption{Upper: Distribution of clump luminosity (Sect.~\ref{sec_lumical}) as a function of $t$ for LMC clumps (blue dots) and the MW sample restricted to distances greater than 3~kpc (gray dots). Dashed lines represent the empirical total luminosity ($L_{\rm tot}$), along with individual contributions from stellar luminosity ($L_{\rm tot}\dag$) and accretion luminosity ($L_{\rm acc}$) as functions of $t$, derived from the full ATLASGAL sample by \citet{2025arXiv251025436L}. Solid purple lines show the sum of the empirical $L_{\text{tot}}$ and an additional $100\,L_{\odot}$. Lower: Same as the upper panel but for the SMC sample, plotted alongside the full MW sample for comparison.}
    \label{fig_lumit}
\end{figure}

\section{Milky Way analogs of Magellanic Cloud clumps} \label{sec:calibration}
In this section, we compare the dust temperature distributions (Sect.~\ref{sec_TdustDistr}), luminosity evolutionary trends (Sect.~\ref{subsec:lumi_evol}), and spatial clustering characteristics (Sect.~\ref{sec_qon}) to evaluate whether the Magellanic Cloud (MC; especially the LMC) clumps serve as suitable analogs for Milky Way (MW) clumps, and if so, which specific subset of the MC clumps provides the closest match.

\subsection{Dust temperature number distribution}\label{sec_TdustDistr}
Here, we compare the $T_{\rm dust}$ number distributions among the LMC, SMC, and MW clump populations. A clear statistical similarity is found, suggesting that these clumps share a common heating mechanism driven by stellar feedback during the star formation process. Consequently, this allows us to directly apply the dust temperature-to-timeline mapping calibrated from MW clumps to the MC clump samples in Sect.~\ref{subsec:lumi_evol}.

\subsubsection{Dust temperature of Large Magellanic Cloud clumps}\label{sec_TdustDistrlmc}
The upper-left panel of Figure~\ref{fig:TdustDistr} shows the number distribution of $T_{\rm dust}$ for LMC clumps across different subgroups (Sect.~\ref{sec_lmcclump}), alongside that of MW clumps (marked as $G_{\rm A}$) for comparison. At the high-temperature end ($T_{\rm dust} > 20$~K), the distributions of LMC $G_1$ clumps and MW clumps are highly comparable in profile shape. Subgroup $G_3$ populates the low-temperature tail of the LMC distribution ($T_{\rm dust} < 13$~K, indicated by the purple vertical line in Figure~\ref{fig:TdustDistr}), while subgroup $G_2$ occupies the intermediate-temperature range ($13 < T_{\rm dust} < 20$~K). The distribution of the combined set $G_1 \cup G_2$ nearly reproduces that of the MW clumps for $T_{\rm dust} > 13$~K, while exhibiting a slight enhancement between $7 < T_{\rm dust} < 13$~K.

For the temperature range $T_{\rm dust} > 13$~K, the number distribution of $G_1 \cup G_2$ is remarkably similar to that of $G_{\rm A}$ when the latter is confined to samples exceeding various distance thresholds (upper-right panel of Figure~\ref{fig:TdustDistr}). We apply a Kolmogorov--Smirnov (K-S)\footnote{K-S tests are calculated using the \texttt{scipy.stats} module in \texttt{Python}.} test between $G_1 \cup G_2$ and $G_{\rm A}$ partitioned into different distance bins. The resulting $p$-values are mostly larger than 0.05 (lower-left panel of Figure~\ref{fig:TdustDistr}), implying that we cannot reject the null hypothesis that they obey the same underlying parent distribution. This statistical agreement indicates that the dust temperature distribution of dense clumps is robust against both MW line-of-sight distance variations and cross-galaxy environmental differences.

We thus suggest that $G_1 \cup G_2$ represents the dense clump sample in the LMC that is analogous to the population in the MW, sharing a similar evolutionary pattern. Subgroup $G_3$ likely consists of more quiescent, extended, or diffuse gas structures that are systematically absent from the comparison sample. Because the ground-based data reduction pipeline of ATLASGAL automatically filters out extended emission larger than $\sim 2.5'$, it selects exclusively for compact clumps with concentrated dust emission \citep{2009A&A...504..415S}. Note that $2.5'$ at a typical MW distance of 5~kpc physically corresponds to $15''$ at the 50~kpc distance of the LMC, which leaves these structures well within the point-source regime for most \textit{Herschel} bands.

\begin{figure}
    \centering
    \includegraphics[width=0.99\linewidth]{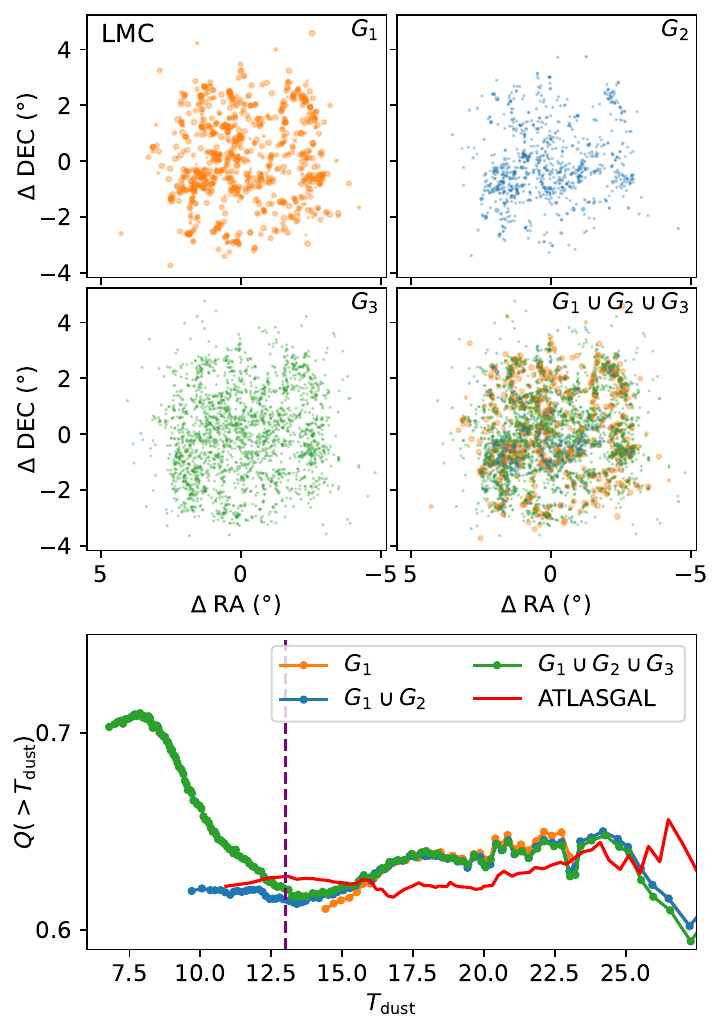}
    \caption{Upper and middle: The face-on spatial distribution (Sect.~\ref{sec_qon}) of 
LMC clumps across different groups (indicated by different colors). Lower: The clustering $Q$ parameters 
calculated for clumps with dust temperatures larger than the given $T_{\rm dust}$ value indicated on the x-axis. The vertical  dashed purple line
marks $T_{\rm dust}=13$~K.}
    \label{fig:LMCdist}
\end{figure}

\subsubsection{Dust temperature of Small Magellanic Cloud clumps}
The $T_{\rm dust}$ number distribution for the SMC (lower-right panel of Figure~\ref{fig:TdustDistr}) is structurally similar to that of the LMC. However, the prominent deficit of $G_2$ sources causes the number distribution between 13 and 25~K to be significantly lower than what would be expected under the assumption of a universal distribution matching the LMC and MW clumps. As demonstrated in the following section (Sect.~\ref{subsec:lumi_evol}), there is no systematic offset in the underlying luminosity versus $T_{\rm dust}$ scaling relations between LMC and SMC clumps. 

One possibility for this deficit is an observational bias caused by the low dust content of the SMC. Because there is less dust shielding, short-wavelength emission from embedded young stars can easily escape without being re-processed by a thick envelope. This makes the warm inner structure more exposed, systematically skewing the fitted dust temperature toward a higher value. Another possibility is a physical pre-heating mechanism. Because the diffuse SMC clumps lack sufficient dust to shield themselves, the ambient interstellar radiation field (ISRF) can easily penetrate the clouds. This warms the youngest clumps to a higher temperature floor and physically shifts them out of the cold $G_2$ regime.

\begin{figure*}
    \centering
    \includegraphics[width=0.99\linewidth]{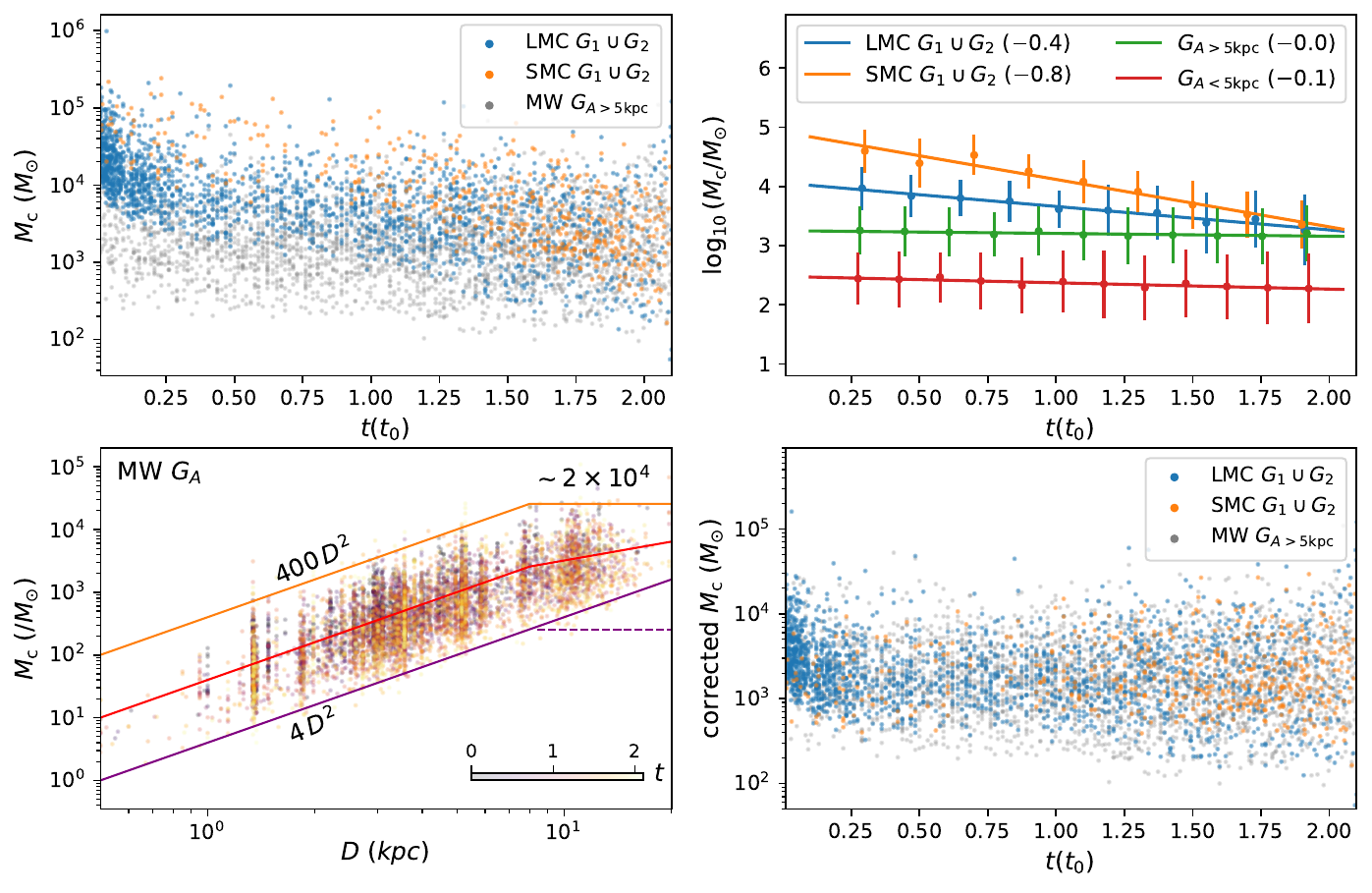}
\caption{Upper left: Clump mass versus $t$ for LMC and SMC dense clumps ($G_1 \cup G_2$), plotted alongside the distant ($D > 5$~kpc) MW sample. Upper right: Mean (dots) and standard deviation (vertical bars) of the logarithmic clump mass evaluated within various temporal bins for each clump population. Solid lines represent linear fits to the binned data, with the resulting slopes ($\beta$, Sect.~\ref{sec:masscor}) indicated in brackets within the legend. Lower left: Mass-distance distribution for the MW $G_{\rm A}$ clumps, where the color scale indicates the normalized evolutionary age $t$. The orange and purple lines enclose the upper and lower observational bounds, while the red line represents the mean of these two boundaries. The purple dashed line (above 8 kpc) marks the lower bound of the undetected, distant lower-mass clumps (Sect.~\ref{sec:clump_scale}). Lower right: Same as the upper right panel, but evaluating the corrected clump masses.}
    \label{fig:masscor}
\end{figure*}

\begin{figure*}
    \centering
    \includegraphics[width=0.88\linewidth]{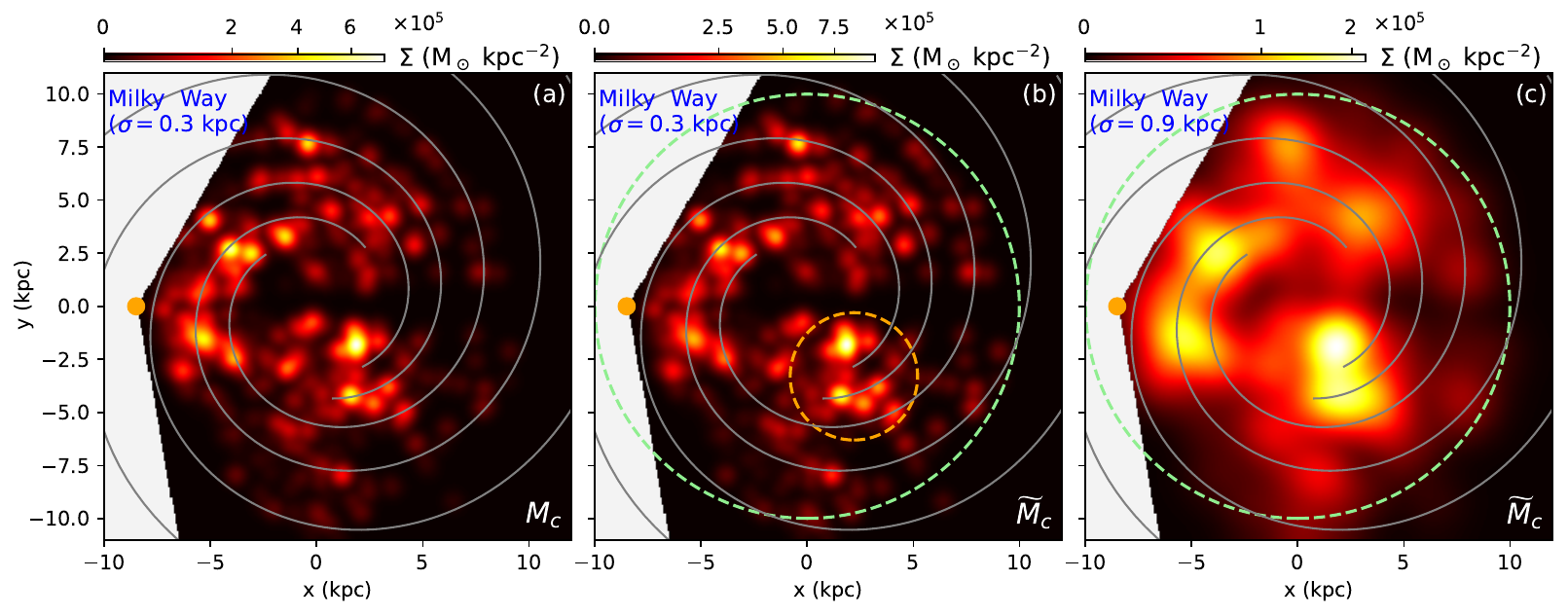}
    \includegraphics[width=0.88\linewidth]{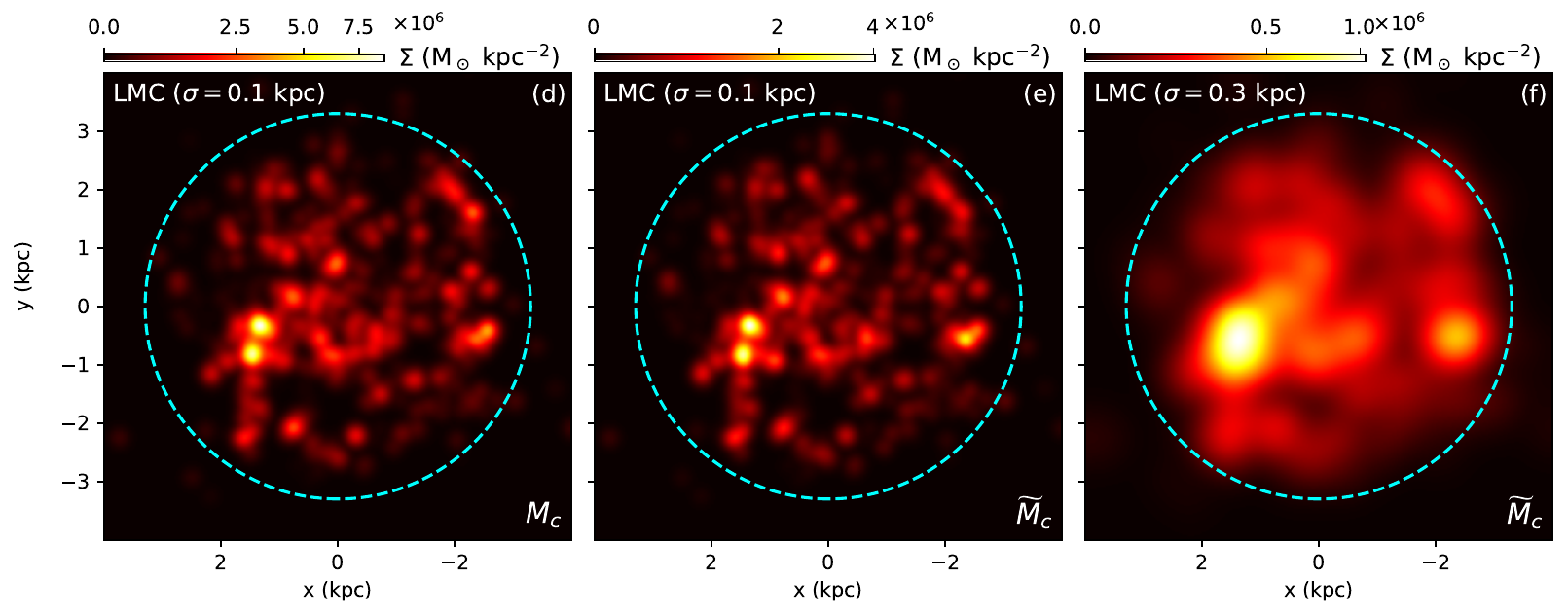}
    \includegraphics[width=0.88\linewidth]{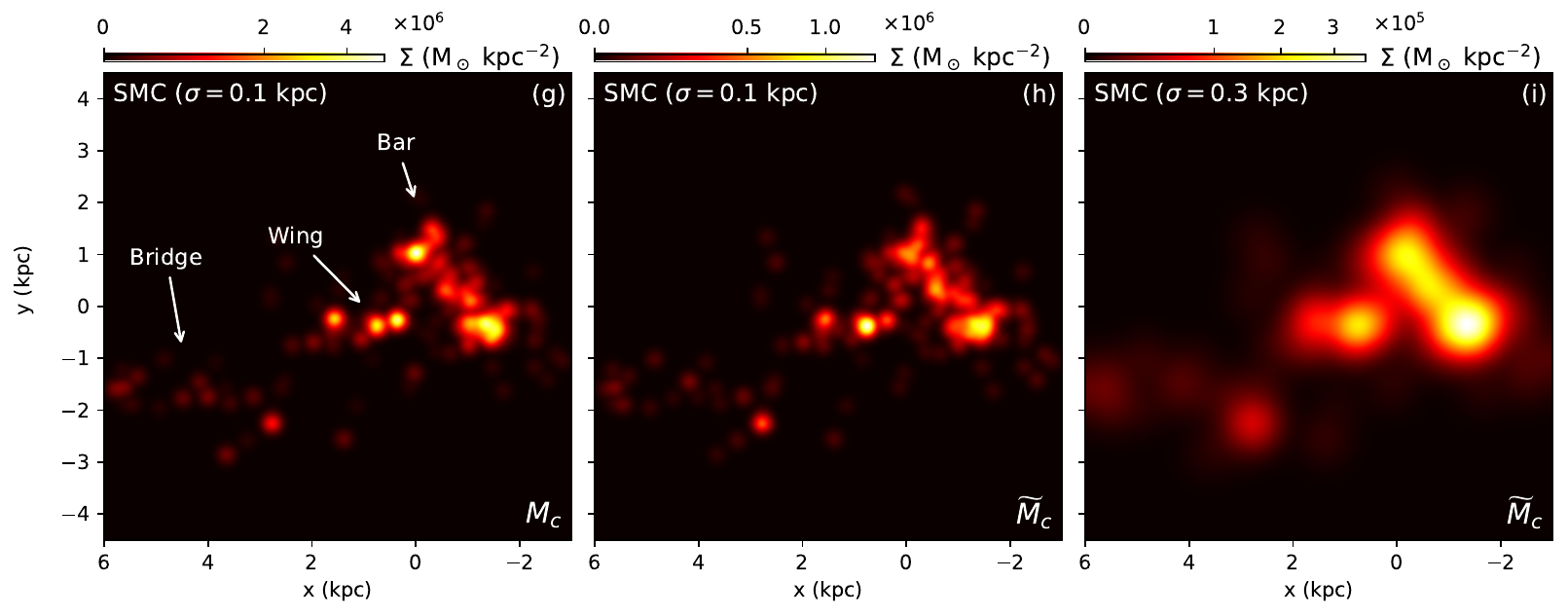}
\caption{Convolved surface density maps (Sect. \ref{sec_render}) for MW (top), LMC (middle), and SMC (bottom) clumps, with the Gaussian standard deviation ($\sigma$) noted in each panel's top-left corner. Column~1 shows uncorrected mass; Columns~2 and 3 show corrected masses (Eqs.~\ref{eq_coMMC} and \ref{eq_coMMW}). Orange and light green dashed circles denote an active region ($R = 3.3$~kpc) and the disk ($R = 10$~kpc) of the MW. Gray lines trace the spiral arm model \citep{2014A&A...569A.125H}, and an orange dot marks the position of the Sun in the first row. The cyan circle marks the LMC disk ($R = 3.3$~kpc). The bar, wing, and bridge of the SMC are marked in panel~g.}
    \label{fig:renderimage}
\end{figure*}

\subsection{Luminosity evolution trend} \label{subsec:lumi_evol}
In Sect.~\ref{sec_TdustDistr}, the LMC $G_1 \cup G_2$ clumps show a remarkable similarity in their $T_{\rm dust}$ number distribution compared to that of the MW clumps. At temperatures below 13~K, the contamination from $G_1$ sources is significant for the LMC clumps, while the $G_2$ sample is entirely missing for the SMC population. Despite these variations, we apply the empirical dust temperature CDF-to-timeline mapping derived from the MW sample \citep{2025arXiv251025436L} directly to the MC clumps to establish their normalized evolutionary age $t$. Using these derived values of $t$, we map the distribution of the MC clumps in the $L_{\rm tot}$--$t$ plane (Figure~\ref{fig_lumit}).

\subsubsection{Large Magellanic Cloud luminosity evolution trend}
Globally, the LMC sample closely follows the empirical evolutionary trend derived for the MW ATLASGAL sample by \citet{2025arXiv251025436L}. This model incorporates an accelerating star formation framework with exponential growth, where the total luminosity is decomposed into an accretion component ($L_{\rm acc} \propto e^t$) tracking the infalling gas reservoir and an embedded stellar component ($L_{\star} \propto e^{3.5t}$) driven by rapidly growing massive young stars inside the clumps. As shown by the dashed lines in the upper panel of Figure~\ref{fig_lumit}, this baseline MW model provides a reasonable first-order description of the LMC data.

However, a discrepancy emerges when analyzing the residuals around this trend. While the full MW ATLASGAL sample exhibits a wide vertical scatter spanning approximately two orders of magnitude, the LMC clumps display a much tighter dispersion of only about one order of magnitude ($\sim 1$~dex). Crucially, these LMC data points are systematically concentrated at the higher-luminosity end of the MW distribution, with the upper boundary well described by the solid purple line in the upper panel of Figure~\ref{fig_lumit}. This difference in scatter and boundary limits is primarily driven by observational selection effects linked to spatial resolution and flux sensitivity.

We find that when the ATLASGAL sample is restricted to distant sources ($D > 3$~kpc, denoted as $G_{\rm A,\ >3\,\rm kpc}$ and plotted as gray dots), this subset exhibits a matching luminosity range and a tight $\sim 1$~dex scatter that mirrors the LMC distribution. In contrast, the full MW sample shows a wider 2~dex scatter (see the lower panel of Figure~\ref{fig_lumit}). Furthermore, this distant MW subset shares a nearly identical $L_{\rm tot}$--$t$ trend line with the LMC sample. An exception occurs in the low-luminosity regime ($t \lesssim 0.5$), where the luminosities of the LMC clumps tend to be systematically higher by about a factor of two; this offset is likely caused by the inclusion of more peripheral, diffuse gas surrounding these colder clumps within the larger \textit{Herschel} beam footprint (Sect. \ref{sec:masscor}). 

Note that this alignment does not imply fundamental physical differences among MW clumps situated at different distances. Rather, a MW clump observed at closer distances benefits from higher linear resolution, allowing its constituent sub-clumps to be resolved and identified individually while its diffuse peripheral emission is filtered out by the ground-based instrument pipeline (Sect. \ref{sec:masscorall}). For the LMC clumps, the coarser physical beam acts similarly to the distant MW observations, smoothing out local substructure and blending peripheral emission. This selection process filters out low-luminosity, isolated sub-clumps and retains only the more massive, high-luminosity blended structures, resulting in the tighter, elevated distribution observed in both the LMC and the distant MW subset.

\subsubsection{Small Magellanic Cloud luminosity evolution trend}
The $L_{\rm tot}$--$t$ distribution for the SMC closely resembles those of the LMC and MW clumps when adopting the $T_{\rm dust}$-to-$t$ mapping calibrated from the MW baseline. This consistency holds despite the anomalous $T_{\rm dust}$ number distribution observed in the SMC clumps, which arises from the prominent deficit of the $G_2$ population. Overall, we suggest that the distant MW ATLASGAL clumps serve as the direct observational analogs to both the LMC and SMC clump samples (see also Sect. \ref{sec:clump_scale}).

\subsection{Clustering characteristics}\label{sec_qon}
Studies utilizing stellar dynamics and Cepheid photometry exhibit a global inclination range of $i \sim 20^\circ$--$40^\circ$ for the LMC disk \citep{2001AJ....122.1807V, 2004ApJ...601..260N}. However, because the dense clumps are distributed within a nearly circular morphology on the sky, no geometric tilt correction is applied to their face-on spatial coordinates (Figure~\ref{fig:LMCdist}). For the MW ATLASGAL sample, source positions within the MW plane are calculated based on their distances and sky coordinates, adopting a Solar distance to the Galactic center of $R_0 = 8.5$~kpc \citep{2015A&A...579A..91W}. This spatial reconstruction enables a direct structural comparison between the LMC and MW clump clustering under a nearly face-on projection. Note that performing a similar projection is challenging for the SMC, as its clumps are primarily concentrated within the highly elongated bar and disrupted tidal stream structures.

We calculate the clustering $Q$ parameters (as described in Appendix~\ref{sec_qmth}) for the LMC and MW clumps on their face-on planes, confined to samples with dust temperatures larger than a given $T_{\rm dust}$ threshold. The lower panel of Figure~\ref{fig:LMCdist} shows the $Q$ parameter as a function of this $T_{\rm dust}$ threshold. The results indicate that $G_{\rm A}$, $G_1$, and $G_2$ have similar $Q$ values of $\sim 0.62$ and exhibit a flat trend as $T_{\rm dust}$ decreases. This behavior reflects the hierarchical, multi-scale clustering architecture of the clumps, which are concentrated along the spiral arms and bars. However, when the $G_3$ subgroup is added to the $G_1 \cup G_2$ combination, the $Q$ parameter increases to a value of $\sim 0.7$ (indicating random or uniform distribution) at the low-$T_{\rm dust}$ end. This increase quantitatively confirms the visual inspection that $G_3$ possesses a more uniform spatial distribution. Therefore, based on these clustering parameters, as well as the distributions of dust temperatures and luminosities (Sect.~\ref{sec_TdustDistr} and Sect.~\ref{subsec:lumi_evol}), we confirm that $G_{\rm A}$ and $G_1 \cup G_2$ (especially those with dust temperatures greater than $13$~K) in the LMC (and probably also the SMC) serve as analogs of MW dense clumps.

\section{Calibration of star formation patterns and rates} \label{sec:lmc_sfr}
Given the similarity between the MC clump distributions and the MW baseline (Sect.~\ref{sec:calibration}), which implies a shared physical pattern for clump-hosted star formation across these systems, it is possible to compare their global star formation rates from a clump-scale perspective. However, considering the different physical resolution scales (which dictate the level of substructure resolved within each clump) and the varying amounts of peripheral mass covered by individual clump footprints, a direct comparison using simple source counts or uncorrected total masses remains challenging. To establish a unified, cross-galaxy star formation framework that overcomes these resolution barriers, we present our core clump mass correction methodology in Sect.~\ref{sec:masscorall}, explore a geometric estimate of the SFR leveraging the hierarchical similarity of their global clump distributions in Sect.~\ref{sec_render}, and quantitatively calculate the integrated star formation rates in Sect.~\ref{sec_qsfr}.

\subsection{Mass correction frameworks}\label{sec:masscorall}
In this section, we suggest that both the observed distant MW clumps and the warmest clumps within the MCs represent fiducial clumps bounded by a spatial scale of $\sim 1$~parsec. On this structural baseline, we correct for the contribution of the peripheral mass in colder, early-phase MC clumps in Sect.~\ref{sec:masscor}, and compensate for the loss of lower-mass clumps in the distant MW data in Sect.~\ref{sec:clump_scale}.

\subsubsection{Magellanic Cloud clump mass correction}\label{sec:masscor}
The upper-left panel of Figure~\ref{fig:masscor} shows the mass distribution of the LMC and SMC dense clumps (confined to $G_1 \cup G_2$) alongside the distant ($D > 5$~kpc) MW sample. In the high-$t$ regime, their distributions are highly comparable. However, as $t$ approaches the lower end, the clump mass remains invariant for the MW sample, whereas that for the LMC increases by a factor of 6, and that for the SMC exhibits an even steeper increase.
To quantify this trend, we calculate the mean and variance of the mass within different $t$ bins for the various samples, as shown in the upper-right panel of Figure~\ref{fig:masscor}. We fit the $\log_{10}(M_{\rm c})$ distribution linearly against $t$:
\begin{equation}
    \log_{10}(M_{\rm c}) = (t-2)\beta + \log_{10}(M_0),
\end{equation}
where $M_0$ represents the characteristic mass of the warmest clumps at the distance boundary (see Sect.~\ref{sec:clump_scale} for the justification of this reference sample choice). For the MW clumps, the mean mass of the distant sample ($D > 5$~kpc) is systematically higher than that of the nearby sample ($D < 5$~kpc), as expected. However, both MW subsets exhibit no variations along $t$. This independence occurs because the 870~$\mu$m flux density is relatively insensitive to variations in dust temperature, resulting in no systematic trend along $t$. Furthermore, closer clumps are more easily resolved into individual sub-clumps, benefiting from the superior physical resolution.

This distance-dependent physical resolution causes the typical observed masses for MW clumps to scale with distance as $M_{\rm c} \propto D^2$ (lower-left panel of Figure~\ref{fig:masscor}), saturating at an upper bound of $\sim 2 \times 10^4~M_{\odot}$ and a mean value of $\sim 2 \times 10^3~M_{\odot}$ for clumps at the largest distances. Crucially, $t$ has no dependence on mass within each distance bin, confirming that $T_{\rm dust}$ is a robust evolution tracer independent of distance (see also Sect.~\ref{sec_TdustDistrlmc}) and mass \citep{2025arXiv251025436L}.

Remarkably, the $M_0$ values for the MC clumps also converge to $\sim 2 \times 10^3~M_{\odot}$, which is comparable to the saturation value observed for the most distant MW clumps. This convergence occurs because the emission of the warmest clumps is concentrated within the shorter-wavelength \textit{Herschel} bands, yielding a physical resolution comparable to that of ATLASGAL for the distant MW sample (see physical interpretation in Sect. \ref{sec:clump_scale}). Conversely, the inflation of the observed masses for MC clumps at lower $t$ is caused by the fact that colder clumps emit predominantly at longer \textit{Herschel} wavelengths; these channels suffer from coarser angular resolution and thus blend in peripheral mass. To subtract this peripheral mass contribution, we apply the following correction:
\begin{equation}
    \widetilde{M}_{\rm c} = 10^{(2-t)\beta} M_{\rm c}, \label{eq_coMMC}
\end{equation}
where $\widetilde{M}_{\rm c}$ denotes the corrected mass for the MC clumps (see the lower-right panel of Figure~\ref{fig:masscor}). By adopting Eq.~(\ref{eq_coMMC}), the total mass of the LMC $G_1 \cup G_2$ population is reduced to 43\% of its uncorrected value.

\subsubsection{Completeness of Milky Way clumps}\label{sec:clump_scale}
The lower-left panel of Figure~\ref{fig:masscor} suggests that below a distance of 8~kpc, the detection of MW dense clumps is complete. Under the angular resolution of ATLASGAL ($19.2''$), this 8~kpc threshold corresponds to a physical spatial scale of $\sim 1$~parsec, which is comparable to the typical separation observed between embedded clusters \citep[e.g.,][]{2024A&A...691A.293Z}. We suggest that 1~parsec represents the natural fiducial boundary of a single, coherent dense clump. A large portion of dense structures with sizes smaller than 1~parsec should be classified as localized sub-clumps nested within a larger parent potential. We discuss the physical interpretation of this structural hierarchy in Sect.~\ref{sec_disclu}.

Assuming that this parsec-scale structural boundary also applies to lower-mass clumps, the distant ($D > 8$~kpc) low-mass population will be systematically filtered out by sensitivity limits due to distance-dependent beam dilution. The purple dashed line in the lower-left panel of Figure~\ref{fig:masscor} represents the inferred lower bound of these undetected lower-mass clumps. Since the clump masses evenly span a logarithmic range of $s_M \sim 1$~dex within each distance bin (where $s_M$ is narrower than the total separation between the upper and lower limits due to mass variance considerations), the missing fraction of the clump mass range at $D > 8$~kpc can be estimated as:
\begin{equation}
    f_{\rm miss} = \frac{2}{s_M}\log_{10}\left(\frac{D}{8\ \rm kpc}\right) = 2\log_{10}\left(\frac{D}{8\ \rm kpc}\right).
\end{equation}
Consequently, to map and analyze the large-scale spatial distribution of MW clumps (Sect.~\ref{sec_render}), we can compensate for this selection effect by scaling the completeness weight of the distant clumps by a factor of:
\begin{equation}
    f_{\rm cor} = \frac{1}{1 - f_{\rm miss}}.
\end{equation}
Note that for convenient reference, $f_{\rm cor}$ is evaluated to be 1.5 and 2.5 at distances of 12~kpc and 16~kpc, respectively. We apply this correction factor directly to individual clumps within each corresponding distance bin via:
\begin{equation}
    \widetilde{M}_{\rm c} = f_{\rm cor} M_{\rm c}. \label{eq_coMMW}
\end{equation}
We remind the reader that while this calculation is executed on individual clumps, it is physically intended as a statistical compensation to account for the population of undetected lower-mass clumps within that distance bin. Accounting for this distance-dependent completeness correction, the total equivalent mass of the entire $G_{\rm A}$ sample is increased to 1.25 times its original, uncorrected value.

\subsection{Rendered distribution and star formation rate estimates}\label{sec_render}
Figure~\ref{fig:renderimage} shows the rendered surface density maps for the MW, LMC, and SMC clump populations, computed using both the uncorrected and corrected clump masses (Eqs.~\ref{eq_coMMC} and \ref{eq_coMMW}). These continuous distributions are constructed by convolving individual clump positions with a 2D Gaussian kernel across a series of spatial scales ($\sigma = 0.1$, $0.3$, and $0.9$~kpc). The resultant maps can be treated as clump-based dense gas maps.

The completeness correction (Sect.~\ref{sec:clump_scale} and Eq.~\ref{eq_coMMW}) for the MW clumps leads to a more balanced surface density distribution between the near and far sides of the Galactic disk (panel b of Figure~\ref{fig:renderimage}) compared to the uncorrected map (panel a of Figure~\ref{fig:renderimage}). Consequently, the corrected distribution is adopted for subsequent analysis.

\subsubsection{Global similarity between dense gas distributions} \label{sec_glosimi}
Before quantitatively estimating the SFR in Sect.~\ref{sec_qsfr}, we visually compare star formation patterns across the MW, LMC, and SMC by exploring the rendered surface density maps. The active disk traced by dense clumps has a radius of $\sim 10$~kpc for the MW, whereas the LMC disk spans a radius of $\sim 3.3$~kpc. The global dense-gas surface density distributions are comparable between the MW and the LMC when rendered with a smoothing scale of $\sigma = 0.3$~kpc and $\sigma = 0.1$~kpc, respectively; this scaling preserves an identical ratio between the physical size of the host system and the chosen rendering smoothing scale (panels b and e of Figure~\ref{fig:renderimage}). This structural alignment suggests that star formation processes in the LMC and MW are similar at the scale of individual dense clumps (Sect.~\ref{sec:calibration}), and that their global spatial distributions of dense clumps share a similar configuration. 

A comparison of the $Q$ parameters quantitatively confirms this global spatial distribution similarity (Sect.~\ref{sec_qon}), though it represents a scale-free statistic. By comparing the rendered maps, we suggest that the ratio between the spatial scales of clump distributions in the MW and the LMC is approximately three, equivalent to the size ratio between their disks. Furthermore, the surface density ratio between the LMC and the MW at these respective smoothing scales is approximately 4, compared to a geometric disk area ratio of 9. This implies that the LMC disk is structurally similar to the MW disk but compressed in physical size by a factor of 3. Because the total number of dense clumps remains of the same order of magnitude despite this spatial compression, our findings demonstrate that the LMC represents a compact, high-density analog of the MW. 

Consequently, the star formation rate of the LMC can be estimated via this structural scaling as:
\begin{equation}
    {\rm SFR}_{\rm LMC} \sim \frac{4}{9} {\rm SFR}_{\rm MW} \sim 0.4~M_\odot\,\rm yr^{-1},
\end{equation}
where the fiducial value of the clump-hosted ${\rm SFR}_{\rm MW}$ is adopted as $1~M_\odot\,\rm yr^{-1}$ \citep[e.g.,][]{2010ApJ...710L..11R}. This estimated ${\rm SFR}_{\rm LMC}$ value is compatible with the literature baseline of $0.2$--$0.8~M_\odot\,\rm yr^{-1}$ evaluated at coarser time resolutions of $\gtrsim 10^7$~yr \citep[e.g.,][]{2009AJ....138.1243H, 2012ApJ...761...42S}. This agreement confirms that the LMC is experiencing a new epoch of ongoing bursty star formation initiated over the past $10^7$~yr \citep{2009AJ....138.1243H}, which is cleanly captured by the dense clump snapshot timescale of $\sim 10^5$~yr.

\subsubsection{Hierarchical distribution of dense clumps} \label{sec_hier}
When the surface density map of the LMC is further smoothed to $\sigma = 0.3$~kpc, it remains globally similar to the MW disk smoothed with a three times larger scale of $\sigma = 0.9$~kpc. Furthermore, these patterns resemble a zoomed-in, 3.3~kpc radius active sub-region of the MW (such as the region enclosed by the orange dashed circle in panel b of Figure~\ref{fig:renderimage}) when rendered with a smoothing scale of $\sigma = 0.3$~kpc. While the SMC clumps appear to follow a similar spatial distribution (bottom panels of Figure~\ref{fig:renderimage}), its highly irregular shape and lack of a well-defined face-on disk geometry prevent us from performing a detailed quantitative analysis of its structure. Nevertheless, this self-similarity suggests that the spatial distribution of dense clumps follows a hierarchical, nested structure that persists across different galaxies with distinct global reference scales.

\subsection{Quantitative calculation of the star formation rate}\label{sec_qsfr}
Here, we estimate the global SFR directly from the corrected clump masses, assuming that the star-forming activities of these galaxies are dominated by those associated with dense clumps. We further assume that on a timescale of $t_{\rm c} \sim 2t_0 \sim 10^{6}$~yr, the mass of a dense clump will be entirely consumed under a star formation efficiency of $\epsilon \sim 10\%$. For a given total mass of clumps, $M_{\rm tot}$, the global SFR is expressed as:
\begin{equation}
    {\rm SFR} \sim \frac{\epsilon M_{\rm tot}}{t_{\rm c}}. \label{eq_clsfrs}
\end{equation}
See Sect. \ref{sec_timescales} for the justification of $t_c$.

The total clump mass of the MW covered by the ATLASGAL survey is $1.1 \times 10^7$~$M_\odot$. For the regions uncovered by the survey, we adopt a mean surface density identical to that of the covered region located at Galactocentric distances between 7~kpc and 10~kpc (Figure~\ref{fig:renderimage}). This extrapolation leads to an estimated uncovered mass of $4 \times 10^5$~$M_\odot$, which is approximately 4\% of the value within the covered region. Combining these components yields a total dense clump mass in the MW of $1.14 \times 10^7$~$M_\odot$, leading to a global SFR of $1.14~M_\odot\,\rm yr^{-1}$, which is compatible with well-adopted literature values.

The value of $M_{\rm tot}$ after applying the correction from Eq.~(\ref{eq_coMMC}) for the LMC is $4 \times 10^6$~$M_\odot$, corresponding to an integrated SFR of $0.4~M_\odot\,\rm yr^{-1}$. This result is compatible with our qualitative estimation discussed in Sect.~\ref{sec_glosimi}. Finally, the total corrected clump mass $M_{\rm tot}$ for the SMC is $10^6$~$M_\odot$, which corresponds to a global SFR of $0.1~M_\odot\,\rm yr^{-1}$. This value is compatible with integrated literature baselines \citep[e.g.,][]{2015MNRAS.449..639R}. An uncertainty of a factor of two for the derived SFR is estimated from the mass correction framework.

\section{Discussion}\label{sec:discu}

\subsection{Self-regulation of maximum clump mass}\label{sec_disclu}
\citet{2024A&A...691A.293Z} argued that currently observed clumps cannot be the direct precursors of typical open clusters, given that only a small fraction of isolated ATLASGAL clumps possess masses exceeding $3\,000~M_{\odot}$. However, in our view, the naturally coherent spatial scale for dense clump complexes is indeed $\sim 1$~parsec (Sect.~\ref{sec:clump_scale}). At this scale, the fiducial mass centers around $2\,000~M_{\odot}$ and the upper mass boundary reaches greater than $10\,000~M_{\odot}$ (see the lower-left panel of Figure~\ref{fig:masscor}). Consequently, a true parent clump that integrates these coherent sub-clumps within its parsec-scale boundary can indeed serve as the direct precursor to a standard open cluster. This framework holds for both the MW and the MCs despite their very different metallicities and environments, and is thus suggested to be universal for the formation of clusters and massive stars.

The core mass function (CMF), especially at its high-mass end, is modulated by both the initial core mass function (ICMF) and its subsequent accretion \citep{2025arXiv251025436L}. To form a very massive star, the ICMF must enable the existence of a very massive seed, which is strongly regulated by the coherent mass of its host clump. Note that, if we treat the initial density field of a coherent structure as a stochastic field with a Hurst parameter of 0.5 (corresponding to Brownian motion), the zero-crossing interval ($L$) along each direction obeys a probability distribution of $\propto L^{-2}$ \citep{2025RAA....25b5020L}. Consequently, the probability of an interval being larger than $L$ scales as $\propto L^{-1}$. The simultaneous probability of having intervals larger than $L$ in all three dimensions is $\propto L^{-3}$. Assuming that the spatial scale tracks the mass reservoir as $L \propto m^{1/3}$, the probability for a seed to possess a mass higher than $m$ scales as $\propto L^{-3} \propto m^{-1}$. This distribution corresponds to a power-law slope of $\alpha = 2$ for the differential ICMF ($p(m) \propto m^{-\alpha}$) at the high-mass end. That is, the stochastic fluctuation of a coherent density field naturally produces a top-heavy seed distribution for the ICMF, and this top-heavy tail is further extended toward a higher cutoff mass through subsequent accretion.

Physically, we interpret a true parent clump (including all its coherent sub-clumps) as a unified structure whose total reservoir co-contributes to the ICMF, particularly shaping its upper mass cutoff. Assuming a linear mapping between the dense core mass and the resulting stellar mass, the maximum stellar mass ($m_{\rm max}$) is expected to scale proportionally with the mass of its parent clump. Adopting a multi-segment Kroupa initial mass function \citep[IMF,][]{2001MNRAS.322..231K} below $0.5~M_\odot$ coupled with a shallow power-law slope of $\alpha = 2.0$ above $0.5~M_\odot$, a massive parent clump at the empirical upper boundary of $\sim 10^4~M_\odot$ (Figure~\ref{fig:masscor}) will form a stellar cluster with a total mass of $1000~M_\odot$ (assuming $\epsilon = 0.1$). 
Integrating this IMF trajectory yields an expected maximum stellar mass of $m_{\rm max} \sim 150~M_\odot$. This derived value sits in agreement with the observed single-star mass ceiling in the local Universe, famously exemplified by the most massive stellar objects hosted within the R136 cluster core in the LMC \citep{2022ApJ...935..162K}. Crucially, such a massive star possesses a short main-sequence lifetime of $\sim 2 \times 10^6$~yr \citep{2015A&A...573A..71K}, which is comparable to the evolutionary lifetime of its parent parsec-scale clump. The subsequent core-collapse supernova explosion of this extreme stellar object will violently disrupt the dynamic coherence of its surrounding environment. Therefore, the maximum coherent mass of an interstellar dense gas structure is fundamentally self-regulated by the feedback of the most massive stars it produces, naturally setting the observed physical upper limits on both the spatial scale ($\sim 1$~parsec) and mass ($\sim 10^4~M_\odot$) of dense clumps.

\subsection{Timescales for nested clumps}\label{sec_timescales}
A constant depletion timescale ($t_0$) may be invalid across different mass regimes. Both compact clumps and low-mass star-forming regions show accelerating star formation, but they operate on distinct timescales \citep{2025arXiv251025436L}. Local low-mass clouds exhibit star formation acceleration over a baseline timescale of $\sim 10^{6}$~yr \citep{2000ApJ...540..255P,2018MNRAS.474.4818C}, which is longer than the typical $\sim 10^5$~yr timescale of massive star formation.

A coherent parent clump is an assembly of multiple sub-clumps. Even though individual sub-clumps collapse on a rapid snapshot timescale of $\sim 10^5$~yr, sequential star formation modulates the integrated timeline of the entire parent clump to $\sim 10^6$~yr. This nested temporal modulation explains why choosing a global consuming timescale of $t_{\rm c} \sim 10^6$~yr in Eq.~\ref{eq_clsfrs} to calculate the SFR is physically valid.

This temporal modulation links high-mass and low-mass star formation within a self-similar framework, reconciling their different timescales through a spatial hierarchy. A low-mass cloud and a massive parsec-scale parent clump both follow a shared mathematical pattern of accelerating star formation over their respective global lifetimes. The critical feature is structural nesting: the parent clump contains a higher density of sub-clumps, each collapsing rapidly on local dynamical timescales. Sequential sub-clump ignition creates localized acceleration events that smoothly integrate into a longer global consumption timeline. Consequently, a parent clump, which produces both massive stars and a bunch of low-mass stars, has an accelerating pattern that is comparable to the longer timescale characteristic of low-mass systems.

\section{Summary}\label{sec:summary}
In this work, we investigate the structural and evolutionary similarities between star and cluster formation patterns at the individual clump scale and the global galaxy scale. We achieve this by comparing the dense clump populations in the Milky Way (MW) detected by the ATLASGAL survey with those detected by the \textit{Herschel} HERITAGE survey in the Magellanic Clouds (MCs, including both the LMC and SMC). Our primary findings are summarized as follows:

\begin{enumerate}
    \item The MW and MC clump populations exhibit consistent dust temperature distributions, mass spectra, luminosity evolutionary trends, and spatial clustering characteristics. This alignment demonstrates that individual dense clumps behave as physical analogs, sharing an invariant localized star formation mechanism despite operating in different global environments.
    \item We establish that the warmest MC clumps and the most distant MW clumps share an identical fiducial parent structure bounded by a natural spatial scale of $\sim 1$~parsec, whereas closer MW clumps are resolved into discrete sub-clumps and colder, early-phase MC clumps suffer from the artificial inclusion of peripheral envelope mass. Crucially, a true parent clump that integrates these coherent sub-clumps within its parsec-scale boundary possesses a sufficient mass reservoir to serve as the direct precursor to a standard open cluster.
    \item The global layout of dense clumps in the LMC and the MW shares a similar pattern when adjusting for the size difference between the two galaxies. This matching pattern arises because the clump distributions look identical as long as we preserve a constant ratio between the physical size of the host galaxy disk and the chosen rendering smoothing scale. While this specific matching ratio could be a coincidence for these two systems, it suggests that the spatial distribution of dense clumps follows a nested, hierarchical structure that may be similar across different galaxies but operating at different intrinsic reference scales.
    \item We calculate the star formation rates ($\sim 0.4~M_\odot\,\rm yr^{-1}$ for the LMC and $\sim 0.1~M_\odot\,\rm yr^{-1}$ for the SMC) based on the corrected clump masses, which are compatible with literature baselines derived at coarser time resolutions of $\gtrsim 10^7$~yr. This close agreement confirms that the LMC is currently experiencing a new epoch of intense, ongoing bursty star formation that began within the past $10^7$~yr, a physical state that is cleanly captured by the instantaneous dense clump snapshot timescale of $<10^6$~yr.
\end{enumerate}

We suggest that high-mass star and cluster formation patterns across different environments are universal. Unbiased observations of larger samples, especially with high enough angular resolutions to resolve the internal structures of dense clumps in different galaxies, will be helpful to confirm this hypothesis.

\begin{acknowledgement}
X.L. acknowledges the support of the Strategic Priority Research Program of the Chinese Academy of Sciences under Grant No. XDB0800303.
\end{acknowledgement}

\bibliography{LMC_SF}
\bibliographystyle{aa} 

\begin{appendix}
\nolinenumbers
\section{Clustering parameter $Q$}\label{sec_qmth}
To quantify the spatial clustering patterns of our sample, we utilize the dimensionless $Q$-parameter introduced by \citet{2004MNRAS.348..589C}. For a given set of $N$ points embedded in a 2-dimensional space, let $\bar{m}$ denote the mean edge length of the Minimum Spanning Tree (MST) connecting the nodes, and let $\bar{s}$ denote the mean separation distance across all unique point pairs. The structural $Q$-parameter is defined as:
\begin{equation}
    Q = \frac{\sqrt{N}\bar{m}}{\bar{s}\sqrt{\pi}}. \label{eq_Qdef}
\end{equation}
This formulation ensures that $Q$ remains scale-independent and does not rely on choosing an arbitrary boundary to enclose the sample. For $N$ points randomly distributed within a 2D unit square, the mean MST edge length scales asymptotically as $\bar{m} \sim 1 / \sqrt{N}$, while the mean pair separation approaches a fixed spatial limit $\bar{s} \sim \text{constant}$. This $N^{-1/2}$ dependency balances the $N^{1/2}$ factor in the numerator, yielding a size-independent baseline value of $Q \sim 0.70$ for a uniform distribution in a square footprint. The factor $1/\sqrt{\pi}$ in Eq.~(\ref{eq_Qdef}) is adopted to maintain compatibility with the original definition by \citet{2004MNRAS.348..589C}.

A single cluster with a monotonically decreasing radial density profile leads to a higher $Q$ value. For instance, in the case where $N-1$ points for a large sample $N$ are co-located at $(0,0)$ and the $N$-th point is situated at $(1,0)$, the mean metrics scale as $\bar{m} \sim 1/N$ and $\bar{s} \sim 1/N$, yielding an unboundedly large parameter value of $Q \sim \sqrt{N}$. Alternatively, for a set of points with a known $Q$, if we split each point into $M$ points (where $M$ is large) at its identical position, $\bar{m}$ will scale down by a factor of $1/M$, while $\bar{s}$ remains invariant. Accounting for the new total sample size $NM$ in the numerator, this configuration results in a suppressed $Q$ parameter that scales down by a factor of $1/\sqrt{M}$ from the original value. At the opposite extreme, if the $N$ points are evenly distributed along a straight line with a constant unit separation of 1, the mean metrics scale as $\bar{m} \sim 1$ and $\bar{s} \sim N$. This linear layout results in a suppressed clustering value of $Q \sim 1/\sqrt{N}$. Consequently, a larger $Q$ value reflects an isolated, centrally concentrated clustering profile, whereas a smaller $Q$ value indicates a hierarchical, multi-scale clustering architecture or a uniform distribution embedded within a lower-dimensional fractal structure.


\end{appendix}

\end{document}